\documentclass[conference]{IEEEtran}
\IEEEoverridecommandlockouts
\usepackage{cite}
\usepackage{amsmath,amssymb,amsfonts}
\usepackage{algorithmic}
\usepackage{graphicx}
\usepackage{textcomp}
\usepackage{xcolor}
\usepackage{amsfonts}
\usepackage{booktabs}
\usepackage{float}
\usepackage{graphicx}
\usepackage{relsize}
\usepackage{grffile}
\usepackage{algorithm}
\usepackage{algorithmic}
\usepackage{xspace}

\usepackage{longtable}
\usepackage{wrapfig}
\usepackage{gensymb}
\usepackage{textcomp}
\usepackage{amsmath}
\usepackage{algorithm}
\usepackage{graphicx}
\usepackage{caption}
\usepackage{listings}
\usepackage{array}
\usepackage{subcaption}
\usepackage{booktabs}
\usepackage{amsmath}
\usepackage{enumitem}
\usepackage{dblfloatfix}
\usepackage{lipsum}
\usepackage{soul}
\usepackage{fancyvrb}
\usepackage{xcolor}
\usepackage{verbatimbox,caption,float,lipsum}
\newfloat{Code}
\def\arraystretch{1.25}
\usepackage[utf8]{inputenc}

\begin{document}
\title{Secure and Efficient Privacy-preserving Authentication Scheme using Cuckoo Filter in Remote Patient Monitoring Network}

\author{\IEEEauthorblockN{Shafika Showkat Moni\IEEEauthorrefmark{1}\textsuperscript{\textsection}, Deepti Gupta\IEEEauthorrefmark{2}\textsuperscript{\textsection}}
\IEEEauthorblockA{\IEEEauthorrefmark{1}Department of Computer Science,
Montclair State University,
Montclair, New Jersey, USA
\\\IEEEauthorrefmark{2}{Department of Computer Science},
{University of Texas at San Antonio},
San Antonio, Texas, USA }
\IEEEauthorrefmark{1}monis@montclair.edu, 
\IEEEauthorrefmark{2}deepti.mrt@gmail.com}

\maketitle
\begingroup\renewcommand\thefootnote{\textsection}
\footnotetext{Equal Contribution}
\endgroup

\begin{abstract}
%With ubiquitous smart medical devices and systems, a large amount of data associated with them is at prime risk from malicious entities (e.g., users, devices, applications) in these systems. Pseudonym, digital signature, and authenticated key exchange (AKE) protocols are used for Internet for Medical Things (IoMT) to develop a secure authorization and preserving privacy of data among patient, smart devices and health practitioners. However, traditional authentication protocols for IoMT devices face overhead challenges due to maintain a large set of key-pairs or pseudonyms results on the hospital cloud server. In this research work, we identify this research gap and introduce a lightweight privacy-preserving authentication scheme using cuckoo filter for Remote Patient Monitoring (RPM) network.

%%%%%%%%%%%%%%%%%%%%%%%%%%%%%%%%%%%%%%%%%%%%%%%%%%%%%%%%%%%%%%%%%%%%%%%%%%%%%%%%%%%%%%%%%%%%%%%%%%%%%%%%%%%%%%

%%%%Abstract by moni%%%%%%%%%%%%%

%%%%%%%%%%%%%%%%%%%%%%%%%%%%%%%%%%%%%%%%%%%%%%%55

With the ubiquitous advancement in smart medical devices and systems, the potential of Remote Patient Monitoring (RPM) network is evolving in modern healthcare systems. The medical professionals (doctors, nurses, or medical experts) can access vitals and sensitive physiological information about the patients and provide proper treatment to improve the quality of life through the RPM network. However, the wireless nature of communication in the RPM network makes it challenging to design an efficient mechanism for secure communication. Many authentication schemes have been proposed in recent years to ensure the security of the RPM network. Pseudonym, digital signature, and Authenticated Key Exchange (AKE) protocols are used for the Internet of Medical Things (IoMT) to develop secure authorization and privacy-preserving communication. However, traditional authentication protocols face overhead challenges due to maintaining a large set of key-pairs or pseudonyms results on the hospital cloud server. In this research work, we identify this research gap and propose a novel secure and efficient privacy-preserving authentication scheme using cuckoo filters for the RPM network. The use of cuckoo filters in our proposed scheme provides an efficient way for mutual anonymous authentication and a secret shared key establishment process between medical professionals and patients. Moreover, we identify the misbehaving sensor nodes using a correlation-based anomaly detection model to establish secure communication. The security analysis and formal security validation using SPAN and AVISPA tools show the robustness of our proposed scheme against message modification attacks, replay attacks, and man-in-the-middle attacks.

%DG: write some lines about how would we analysis the performance 
\end{abstract}
\section{Introduction}
The Internet of Medical Things (IoMT) has become increasingly popular due to improving the quality and accessibility of healthcare services. IoMT devices and their data-driven applications provide various outcomes, including better lifestyle, early disease diagnosis, better quality treatment, and cost-effective Remote Patient Monitoring (RPM) network. During the Covid-19 pandemic, the RPM network has become increasingly popular, especially in minimizing the spread of coronavirus. As the worldwide Coronavirus pandemic continues, more healthcare providers utilize the RPM network to ensure patient and medical professional safety. It is estimated\footnote{https://www.forbes.com/sites/forbestechcouncil/2022/02/14/the-future-of-rpm-pandemic-driven-solutions/?sh=62d6d5411ab7} that over 30 million patients will use the RPM network in the United States by 2024. 

The RPM network consists of smart medical devices, edge devices, communication protocols, patients, medical professionals, and advanced technologies (e.g., cloud computing, edge and fog computing, multi-factor authentication, blockchain, machine learning, and virtual reality). Patient-to-Medical professional (P2M), Device-to-Cloud (D2C), Device-to-Edge (D2E), and Device-to-Device (D2D) communications are widely used in the RPM network, which provide instant care to patients. In the RPM network, the physiological data of patients are collected through sensor nodes and transmitted to the medical server over an open wireless channel. The medical professionals access and analyze the patient's medical data and provide a proper diagnosis. Although the advantages of the RPM network are significant, malicious entities can create various issues, including modifying, intercepting, and replaying the messages (vital readings, prescription) in the RPM network due to its wireless nature of communication. Since these messages contain sensitive and life-critical information, security and privacy are crucial in the RPM network.

In the past, various authentication schemes \cite{hayajneh2016secure, alzahrani2020secure, soni2021privacy, shuai2020efficient} have been developed to secure the RPM network. 
Researchers discuss some of the vulnerabilities that exist among patients, smart devices, and medical professionals, however, the RPM network is still facing challenges to develop a proper authentication scheme. More recently the US National Institute of Standards and Technology (NIST) published a report~\cite{cawthra1800securing} on the RPM ecosystem, which highlights possible security and privacy solutions to build a secure RPM network.

Therefore, a patient needs to check the authenticity of a received message as well as the legitimacy of the medical professionals to accept the message. To mitigate this research gap, we focus on developing a secure and efficient privacy-preserving authentication scheme for patient-to-medical professional communication in this research. 

%%In the past, to secure the identity of user and ensure privacy, differential privacy \cite{dwork2014algorithmic} and data masking techniques \cite{vasilomanolakis2015security} (pseudonymize \cite{kobsa2003privacy}, anonymize \cite{truta2006privacy}) are used widely. In a pseudonymize based approach

The main contributions of this paper are as follows.
\begin{itemize}
    \item We identify the issue of proper authentication mechanism in the Remote Patient Monitoring (RPM) network and present a system model for RPM. 
    \item We propose a novel privacy-preserving authentication mechanism for the RPM network. In our proposed scheme, medical professionals and patients anonymously authenticate each other and establish a shared secret key for secure and efficient communication in the RPM network.
     \item We introduce the concept of the cuckoo filter to reduce the storage, computation, and communication overhead associated with authentication in the RPM network.
    \item We show the proper mechanism for identifying misbehaving sensor nodes, which secure mutual communication.
    \item We also present the security analysis and formal security validation using SPAN and AVISPA tools as a proof-of-concept. 
\end{itemize}

The remainder of this paper is organized as follows. Section~\ref{related} presents the literature review on authentication protocols in healthcare and other domains. We discuss the security requirements in the RPM network in Section~\ref{threat}. Section~\ref{system} presents the proposed system model for the RPM network. In Section~\ref{cuckoo}, we explain the cuckoo filter that makes efficient insertion, deletion, and lookup operations. Section~\ref{proposed} defines the proposed privacy-preserving authentication scheme based on the cuckoo filter and also presents a novel approach to identifying misbehaving sensor nodes. We analyze the security of our proposed authentication scheme and also provide a formal security validation in Section~\ref{security}. Conclusion and future work are discussed in Section~\ref{conclusion}.

\begin{figure*}[t]
\centering
\includegraphics[width=0.8\textwidth, height=.36\textheight]{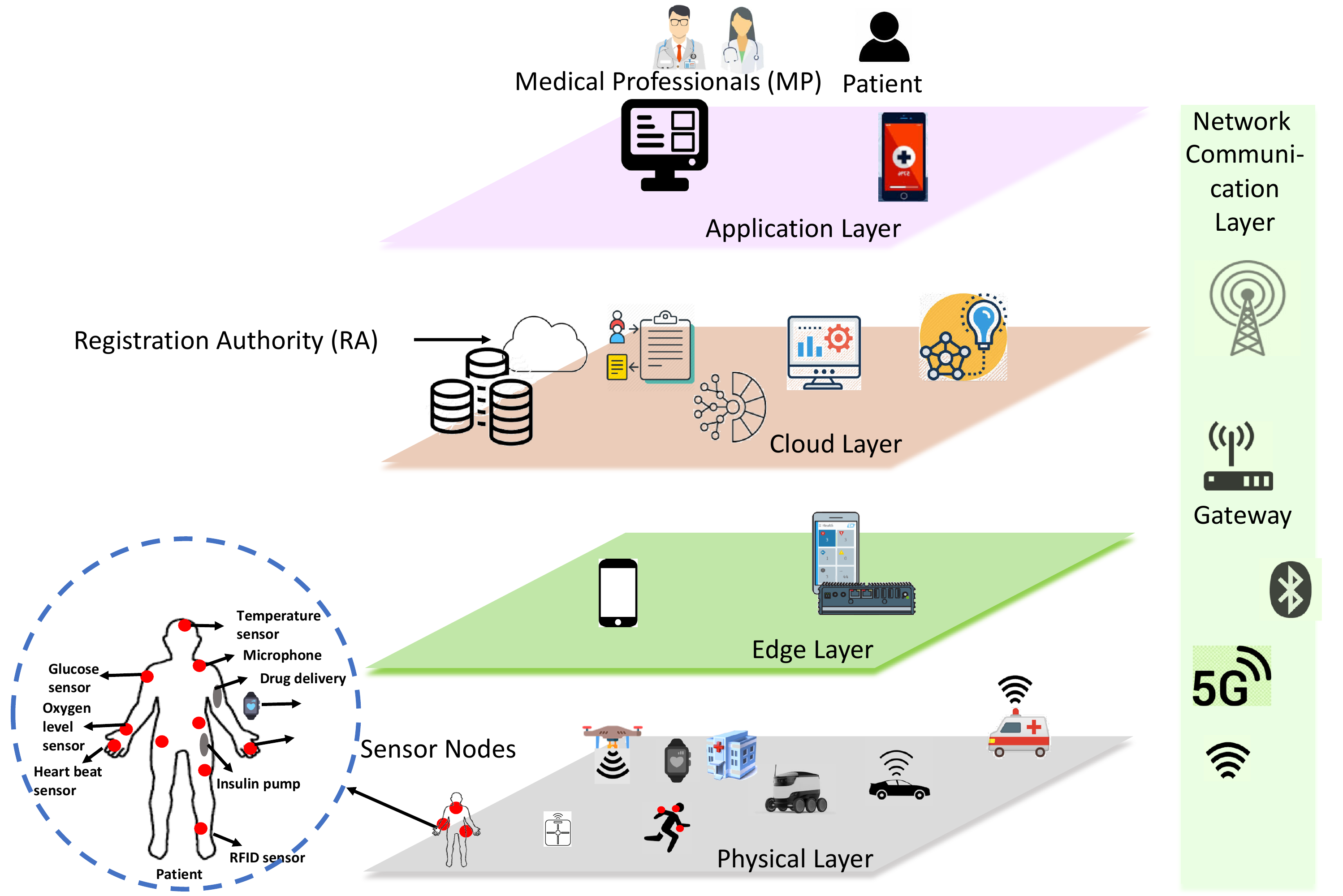}
\centering
\caption{System Model for Remote Patient Monitoring}
\label{system model}
\end{figure*}

\section{Related Work}
\label{related}
This section discusses the related work on secure and efficient authentication protocols for patients, smart devices, and medical professionals. Later, we also present some literature reviews on authentication protocols in different domains, including vehicle and critical infrastructure networks, etc.  

Alzahrani et al.~\cite{alzahrani2020secure} proposed an efficient authentication protocol for RPM that helps to solve many drawbacks, including session key compromise, stolen smart card attacks, and user impersonation attacks. Hayajneh et al.~\cite{hayajneh2016secure} presented a lightweight public-key-based authentication protocol for wireless Medical Sensor Networks (MSNs) to secure the medical network. Soleymani et al.~\cite{soleymani2022privacy} identified the issue in an edge-enabled Internet of Medical Things (IoMT) system and proposed an authentication scheme consisting of digital signature and Authenticated Key Exchange (AKE) protocol for the smart healthcare system. Soni et al.~\cite{soni2021laka} proposed a secure and lightweight health authentication and key agreement protocol using low-cost operations. They also evaluated the proposed protocol against various security attacks and identified that it comparatively takes less execution cost, computation time, and power consumption.

Ali et al.~\cite{ali2020securing} proposed an improved version of a temporal credential-based anonymous lightweight authentication scheme (TCALAS), which is called iTCALAS for the Internet of Drones (IoD) network. The proposed scheme, while maintaining the lightweight, provides security against many known attacks, including traceability and stolen verifier. Shahidinejad et al.~\cite{shahidinejad2021light} introduced a lightweight authentication protocol for IoT devices named Light-Edge using a three-layer scheme. This three-layer scheme contains an IoT device layer, a trust center at the edge layer, and cloud service providers. This proposed scheme works against attack resistance, communication cost, and time cost. Moni and Manivannan~\cite{moni2022lightweight} proposed a cuckoo filter-based lightweight authentication protocol for Vehicular Ad-hoc Networks (VANETs) by solving the overhead issue in traditional Certificate Revocation List (CRL). A CRL is a list of digital certificates that have been revoked by the issuing certificate authority (CA) before their actual or assigned expiration date and is widely used to store the revoked vehicles in a pseudonym-based approach. Grashöfer et al.~\cite{grashofer2018towards} proposed cuckoo filters to improve space efficiency for set membership testing in Network Security Monitoring and presented the example of threat intelligence matching. Yang et al.~\cite{yang2021blockchain} proposed a Blockchain-based authentication scheme to solve security issues. This proposed approach combines the Blockchain technique and the modular square root algorithm to achieve an effective authentication process. In addition, several security models for protecting IoT devices are discussed in~\cite{gupta2020access, gupta2020learner, moni2020scalable, gupta2021game, gupta2021future, aslan2021intelligent, ozkan2021comprehensive, gupta2021detecting, gupta2021hierarchical}.

Prior research has introduced various authentication protocols for medical and other domains, which have been developed to protect from malicious attackers. However, a proper data structure is still missing to reduce the storage, computation, and communication overhead associated with authentication in the healthcare domain. Our proposed privacy-preserving authentication scheme based on the cuckoo filter addresses this issue and provides secure communication between patients and medical professionals in the healthcare environment. The exploration of cuckoo filters makes our scheme lightweight by mitigating the overhead related to authentication in the RPM network. Both the patients and medical professionals store and check the cuckoo filters to authenticate each other and establish a secret shared key for secure and fast communication in the RPM network. In addition, an anomaly detection model is provided in this research to identify misbehaving sensor nodes.
\section{Security Requirements in RPM Network}
\label{threat}
In this section, we discuss the security requirements of the RPM network as follows:
\begin{itemize}
\item Privacy preservation: Ensuring privacy of users (patients and medical professionals) is crucial in the RPM network. Attackers can track the real identity of the users and they can get a lot of sensitive information from the users, especially personal data of the patient. Therefore, the real identity of the patients and medical professionals should be preserved in the RPM network.
\item Mutual authentication: Authentication ensures the legitimacy of the users that participate in the RPM network. Mutual authentication between patient and medical professional makes the communication in the RPM network secure and efficient.
%ISSUE
\item Traceability and Revocation: Although privacy of the medical professionals and patients is important, trusted authority should be able to track the real identity of the medical professionals and patients in case of any disputes. The trusted authority also should be able to revoke the misbehaving medical professionals and patients from the RPM network.
\item Resistance to replay attack: In case of replay attack, the attackers repeat the transmitted messages several times to delay the communication and create confusion in the RPM network. Therefore, the receivers in the RPM network must be able to resist the replay attack for secure communication.
\item Resistance to message modification attack: The attacker may modify the contents of the transmitted messages and breach the security in the RPM network. The medical professionals and patients should be able to determine the message modification attack in the RPM network.  

\end{itemize}

\section{System Model}
\label{system}

Our system model depicted in Fig.~\ref{system model} has five entities: Registration Authority (RA), Medical Professionals (MP), Patients (P), Sensor nodes, and Edge devices. The notations used in this paper are listed in Table~\ref{notations_table}.\\
\begin{table}[!b]
\caption{Notation and Description}
\begin{tabular}{ |p{2.4cm}|p{5.5cm}| }

 \hline
 Notation & Description\\
 \hline
 RA & Registration Authority  \\

 MP & Medical Professional\\
  P & Patient\\
  RPM & Remote Patient Monitoring\\

 ${E}$ & RSA-1024 bit encryption algorithm\\
  ${PU_{RA}}$, ${PR_{RA}}$ & Public and Private  Keys of the RA\\
 ${PU_{MP}}$, ${PR_{Mp}}$  & Public and Private Keys of MP\\
 ${PU_{P}}$, ${PR_{P}}$  & Public and Private Keys of P\\

${PID_{MP}}$& Pseudo-ID of Medical Professional\\
${PID_{P}}$& Pseudo-ID of Patient\\

${t_{s}}$ & Message generation timestamp \\
$PCF$& Positive Cuckoo Filter\\
$NCF$& Negative Cuckoo Filter\\

\hline

\end{tabular}
\label{notations_table}
\end{table}

\begin{itemize}
\item  \textbf{Registration Authority (RA):} Registration authority (RA) is a trusted entity, and RA generates its own public and private key pairs (${PU_{RA}}$, ${PR_{RA}}$). Each entity knows the public key of RA ${PU_{RA}}$, while the private key ${PR_{RA}}$ is kept secret. In the RPM network, each medical professional and patient registers with the RA. As part of the registration process, they provide the necessary information, such as name, address, phone number, real identity, etc. RA is responsible for generating public and private key pairs for every registered medical professional (${PU_{MP}}$, ${PR_{MP}}$) and patient (${PU_{P}}$, ${PR_{P}}$). RA also distributes required credentials (i.e, user ID and password) to the medical professionals and patients to get access to the RPM network securely.

\item\textbf{Medical Professionals (MP):} 
RA generates pseudo-ID ${PID_{MP}}$ for each registered medical professional (doctors, nurses, or medical experts). The MPs receive health-related information from registered patients. They verify the authenticity of the patient as well as the message received from the patient for secure communication. After successful verification, the medical professional provides proper treatment to the corresponding patient.
\item\textbf{Patients:} Each patient gets a pseudo-ID (${PID_{P}}$) from the RA. They use ${PID_{P}}$ instead of their real identity for communication in the RPM network. After registration, the patient submits his/her medical diagnosis reports to the RA. Next, the RA selects appropriate sensor nodes for the patient. The patient also registers the implanted or wearable sensor nodes in his/her body along with the edge device. The edge device collects data from the sensor nodes and forwards the patient's health information to the authorized medical professional. 
\item\textbf{Sensor nodes:} In the RPM network, the sensor nodes can sense and monitor body temperature, heartbeat, pulse, oxygen level in the blood, blood pressure, etc. The patient sends this captured information to the corresponding medical professional.     
\item \textbf{Edge devices:} The edge device ( smartphone, notebook, computer) has enough storage and computational capability. The edge device forwards the information to the patients and medical professional respectively.

\end{itemize}

\section{Cuckoo Filter}
\label{cuckoo}

A cuckoo filter~\cite{fan2014cuckoo} is a probabilistic data structure that uses a cuckoo hash table to perform the set membership test faster. Cuckoo filter stores the fingerprint 
$F(x)$ of an element $x$ instead of storing the original element itself in the cuckoo hash table. The cuckoo filter is formed by an array of $m$ buckets where each bucket has a $n$ number of entries. Each element $x$ has two candidate buckets $i$ and $j$ determined by two hash functions as follows:\\

$i=$ $H_{1}(x)$$=hash(x)$ $mod$ $m$\\
$j=$ $H_{2}(x)$$=(H_{1}(x)$ $\oplus$ $hash(F(x)))$ $ mod$ $m$\\ 

 \begin{figure}[!h] 
  \centering   
  \includegraphics[scale=0.5]{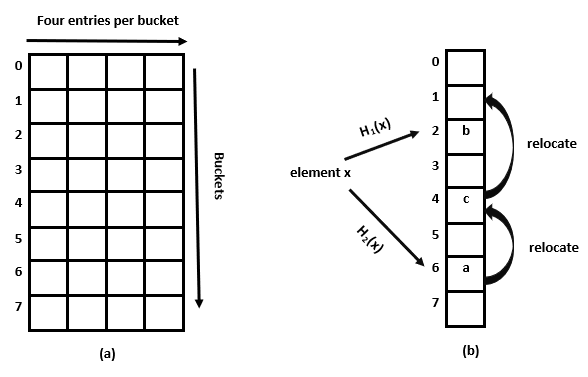}
  \caption{(a) A cuckoo filter (b) Insertion operation.}\label{cuckoo_filter}
\end{figure}

Fig.~\ref{cuckoo_filter}(a) displays a cuckoo filter with eight buckets ($m$ = 8), where each bucket has four entries ($n$ = 4). The cuckoo filter uses two hash functions ($H_{1}(x)$, $H_{2}(x)$) to find the candidate buckets for an element $x$ in the filter. It places $F(x)$ in any of the two empty candidate buckets. The cuckoo filter randomly selects one of the buckets if both candidates are occupied and replaces the existing fingerprint $F(y)$ of the element $y$ in the bucket with $F(x)$. Then inserts the $F(y)$ in its alternate candidate bucket. The alternate bucket location can be found by XORing the 
hash of the fingerprint of the element and the current bucket location. If none of the buckets are empty, a fingerprint in one of the buckets is displaced. This process is repeated until all displaced fingerprints are re-inserted in the cuckoo filter. Fig.~\ref{cuckoo_filter}(b) shows that when inserting an element $x$, it finds that both the candidate buckets (bucket numbers 2 and 6) are occupied. Then selects bucket number 6 randomly and replaces the existing element ($``a"$) in the bucket with $F(x)$. By displacing the existing item ($``c"$) in bucket number 4, the cuckoo filter relocates the kicked-out element ($``a"$) in its alternate candidate bucket number 4. 
It then re-inserts the kicked-out item ($``c"$) into its alternate candidate bucket 1.
The cuckoo filter computes the fingerprint of the element $F(x)$ to look up an element $x$. Then, find the candidate buckets using $H_{1}(x)$ and $H_{2}(x)$, and check the fingerprint $F(x)$ against the fingerprint stored in these buckets. The cuckoo filter returns a positive result whenever a fingerprint matches $F(x)$. If none matches, it returns a negative one. An element $x$'s fingerprint is first looked up in the cuckoo filter using the lookup operation to delete it. Next, delete $F(x)$ when it is found in one of the buckets. Both lookup and deletion operations require a time complexity of $O(1)$ for the cuckoo filter.

\section{Proposed Privacy-preserving Authentication Scheme}
\label{proposed}

In this section, we describe the proposed privacy-preserving authentication scheme that leverages cuckoo filters to authenticate medical professionals and patients in the RPM network efficiently. After successful registration, both medical professionals and patients get pseudo-ID from the RA. They use their pseudo-IDs instead of their real identities for communication in the RPM network. Only RA knows the real identities of the medical professionals and patients, both entities work as anonymous. RA can revoke the misbehaving medical professionals and patients from the RPM network in case of any disputes. In our scheme, RA initializes two cuckoo filters: positive cuckoo and negative cuckoo filters for both medical professionals and patients. RA inserts the fingerprint of pseudo-ID of every registered and valid medical professional in a positive cuckoo filter $({PCF_{MP}})$ and that of misbehaving medical professionals in a negative cuckoo filter $({NCF_{MP}})$. For each registered and valid patient, RA inserts the fingerprints of the patient in a positive cuckoo filter $({PCF_{P}})$ and that of the misbehaving patient in a negative cuckoo filter $({NCF_{P}})$. RA signs both the positive and negative cuckoo filters $Sig_{{RA}{({PCF_{MP}}, {NCF_{MP}})}}$ and $Sig_{{RA}{({PCF_{P}}, {NCF_{P}})}}$, and broadcasts them periodically. Both the medical professionals and patients use the latest cuckoo filters to authenticate each other.  

%\begin{figure*}[t]
%\centering
%\includegraphics[width=0.9\textwidth, height=.25\textheight]{RPM.pdf}
%\centering
%\caption{Remote Patient Monitoring Use Case}
%\label{fig:Usecase}
%\end{figure*}

\subsection{Construction of cuckoo filters for Medical Professional Authentication}
The insert, delete, and search operation in positive cuckoo filter $({PCF_{MP}})$ and negative cuckoo filter $({NCF_{MP}})$ of medical professionals occur as follows:   

\begin{itemize}

\item When a Medical Professional registers with the RA: When a medical professional $MP_{i}$ registers with the RA, he/she gets a pseudo ID $PID_{MP_{i}}$ from the RA for communication in the RPM network. Next, the RA inserts the pseudo ID of the registered medical professional $PID_{MP_{i}}$ in a positive cuckoo filter $({PCF_{MP}})$.    
\item When the RA finds a misbehaving Medical Professional: 
In some cases, an authenticated medical professional $MP_{i}$ may prescribe incorrect or improper treatment to the patient. RA can track the real identity of the misbehaving medical professional $MP_{i}$ using an underlying algorithm and revoke him/her from the RPM network. Next, the RA deletes the pseudo ID $PID_{MP_{i}}$ of the revoked medical professional from the positive cuckoo filter $({PCF_{MP}})$ and stores it in the negative cuckoo filter $({NCF_{MP}})$. The RA also informs the victim patient $P_{i}$ about the misbehaving medical professional $MP_{i}$ so that the patient can no longer communicate with him or her in the future.

\item When a Medical Professional leaves his/her practice: Before leaving his/her professional practice, the medical professional $MP_{i}$ is responsible for notifying the RA. Next, the RA verifies the credential of the corresponding medical professional $MP_{i}$ and revokes him/her from the RPM network. Afterward, the RA removes the pseudo ID of the medical professional $PID_{MP_{i}}$ from both positive and negative cuckoo filters. In addition, the RA informs the corresponding patient under the observation of that medical professional $MP_{i}$ and assigns a new medical professional $MP_{j}$ for him/her.

\end{itemize}

\subsection{Construction of cuckoo filters for Patient Authentication}

\begin{itemize}
\item When a Patient registers with the RA: Upon successful registration, the RA assigns a pseudo ID $PID_{P_{i}}$ to the registered patient $P_{i}$ for communication in the RPM network. After that, the RA inserts the corresponding pseudo ID $PID_{P_{i}}$ in the positive cuckoo filter $({PCF_{P}})$ for more efficient authentication.  
\item When RA finds a Misbehaving Patient: In general, the RPM ecosystem consists of smart health devices, cloud, and various protocols. In our research, we consider that the RPM ecosystem consists of smart health and smart home devices to develop a robust anomaly detection model, which identifies misbehaving sensor nodes. Based on misbehaving sensor node, RA deletes the corresponding pseudo ID $PID_{P_{i}}$ from the positive cuckoo filter $({PCF_{P}})$ and inserts it into the negative cuckoo filter $({NCF_{P}})$. 
\item When a Patient no longer needs treatment: When a patient $P_{i}$ is fully recovered or healed, he /she may no longer stay in the RPM network. The RA can notify the patient about this, or the patient may request the RA to leave the RPM network. In such case, the RA deletes the pseudo ID $PID_{P_{i}}$ of the corresponding patient $P_{i}$ from both the positive cuckoo filter $({PCF_{P}})$ and negative cuckoo $({NCF_{P}})$ filter.

\end{itemize}

\begin{algorithm}[h]
\caption{ Lookup Operation $(PID_{x}) $}
\label{algo1}
\begin{algorithmic}[1]

\STATE {$\zeta=$$fingerprint$ $({PID_{x}})$
\STATE $i=$ $H{{(PID_{x})}}$ $mod$ $m$, where {$m$ is the number of buckets}} 
\STATE {$j=$ $i$ $\oplus$ $H(\zeta)$ $mod$ $m$}
 \IF {$\zeta$ $\in$ $i$ $\lor$ $j$ in $PCF$ and {$\zeta$ $\notin$ $i$ $\land$ $j$ in $NCF$} }
\STATE {${PID_{x}}$ is considered valid}

 \ELSE 
 \STATE {Sends a message to the RA for verification}
\ENDIF
 \IF {$\zeta$ $\notin$ $i$ $\land$ $j$ in $PCF$ and $\zeta$ $\in$ $i$ $\lor$ $j$ in $NCF$  }

\STATE {${PID_{x}}$ is considered malicious}
\ELSE
\STATE {Waits for updated CFs from the RA}
\ENDIF
\end{algorithmic}
\end{algorithm}

\subsection{Authentication Phase}
After successful registration, the patient submits its valid user name and corresponding password to access the RPM network. The RA checks the validity of the patient using cuckoo filters and its database. Upon successful validation, the RA selects an appropriate medical professional $MP_{i}$ for the patient and sends the pseudo ID  ${PID_{MP_{i}}}$ and public key ${PU_{MP_{i}}}$ of the medical professional by encrypting it with the public key of the patient ${PU_{P_{i}}}$ along with message generation timestamp $t_{s}$. After receiving the message from the RA, the patient decrypts the message using his /her private key ${PR_{P_{i}}}$. Firstly, he/she verifies the freshness of the received message using message generation timestamp $t_{s}$. Then, the patient checks the positive cuckoo filter $({PCF_{MP}})$ and negative cuckoo filter $({NCF_{MP}})$ to verify the authenticity of the medical professional $MP_{i}$. Algorithm~\ref{algo1} presents the lookup operation in positive and negative cuckoo filters in detail. Table~\ref{table1} illustrates the four possible outcomes of the query results of positive and negative cuckoo Filter (CF). The query result of case 1 considers the corresponding $PID$ as valid, whereas that of case 2 verifies the $PID$ as malicious. Case 3 indicates that the cuckoo filters have not been updated. In such a scenario, the patients and medical professionals can wait for the updated cuckoo filters from the RA. Case 4 may occur due to the false positive from the cuckoo filters. If case 4 happens, the patients and medical professionals forward the corresponding $PID$ to the RA for further verification. The RA verifies the authenticity of the corresponding $PID$ using its log table, database, and old cuckoo filters. When the $PID$ is verified as valid, RA deletes the corresponding $PID$ from the negative cuckoo filter (NCF). Otherwise, the RA deletes it from the positive cuckoo filter (PCF). Next, the RA distributes the updated cuckoo filters to the patients and medical professionals. 

\begin{table}[!h]
\caption{Four Possible Outcomes of Cuckoo Filters}
\label{table1}
\centering
\begin{tabular}{||c c c c||} 
\hline 
Case & Result of PCF & Result of NCF & Conclusion \\ 
\hline\hline
1 & True & False & Valid \\ 
 2 & False & True & Malicious \\
 3 & False & False & Waits for updated filters \\
 4 & True & True & Forwards to the RA\\ [1ex] 
 \hline

 \end{tabular}

\end{table}

\begin{table*}[t]
\small
\centering
\caption{Authentication process among patient, RA, and medical professional}
%\begin{tabular}{p{.95\textwidth}}
 
\label{M1}
\renewcommand{\arraystretch}{1}
\begin{tabular}{|p{6.1cm} p{4.5cm} p{6cm}|} \hline
\qquad \qquad \qquad \textbf{Patient $(P_i)$}  & \textbf{Registration Authority $(RA)$}  & \qquad \qquad \textbf{Medical Professional $(MP_{i})$} \\ \hline
\begin{itemize}
    \item Sends username and password to RA 
\end{itemize}
  & & \\
  & \begin{itemize}
  \item Received username 
  and password
   
  \item Starts verification
  \item After successful verification

  \item Sends to $P_{i}$ the following message-
   \end{itemize}
 %\quad E$((PID_{MP_{i}}$,$PU_{MP_{i}}$,$t_{s})$,$PU_{P_{i}})$
  %&\\
    $\xleftarrow{\makebox[4cm]{ E$((PID_{MP_{i}}$,$PU_{MP_{i}}$,$t_{s})$,$PU_{P_{i}})$}}$ &\\
%\underleftarrow{$E$ $((PID_{MP_{i}}$,$PU_{MP_{i}}$, $t_{s})$, $PU_{P_{i}})$} & \\
\begin{itemize}
\item Decrypts received message using $PR_{P_{i}}$ 
\item Checks $t_s$ 
\item Retrieves $PID_{MP_{i}}$ and $PU_{MP_{i}}$ 
\item Looks up into $PCF_{MP}$ and $NCF_{MP}$ to verify $PID_{MP_{i}}$ 
\item After successful verification 
\end{itemize}
\quad (a) Generates prime numbers $\alpha$ and $g$ & & \\

\quad (b) Computes A = $g^x$mod$\alpha$, where x is a & & \\
\qquad secret number & & \\ 
\quad (c) Sends to $MP_i$ the following message-& & \\
%$PID_{P_{i}}, PID_{MP_{i}},E((A,\alpha,g, PU_{P_{i}}, t_{s}), PU_{MP_{i}})$  & & \\
$\xrightarrow{\makebox[6.1cm]{$PID_{P_{i}}, PID_{MP_{i}},E((A,\alpha,g, PU_{P_{i}}, t_{s}), PU_{MP_{i}})$}}$ & &\\
%\underrightarrow{$PID_{P_{i}}, PID_{MP_{i}},E((A,\alpha,g, PU_{P_{i}}, t_{s}), PU_{MP_{i}})$}  

& & \begin{itemize}
\item Decrypts message using $PR_{MP_{i}}$ 
\item Checks $t_s$ 
\item Retrieves $(A,\alpha,g, PU_{P_{i}})$ 
\item Looks up into $PCF_{P}$ and $NCF_{P}$ to verify $PID_{P_{i}}$ 
\item After successful verification
\end{itemize}\\ 
& & \quad (a) Computes B = $g^y$mod$\alpha$, where y\\ 
& & \qquad is a secret number\\
& & \quad (b) Computes $K_{s}$ = $A^y$mod$\alpha$\\  
 
& & \quad (c) Sends to $P_i$ the following message-\\

%& & \quad $PID_{MP_{i}},PID_{P_{i}},E((B,K_{s},t_s), PU_{P_{i}})$\\
& & 
%\quad\underleftarrow{$(PID_{MP_{i}},PID_{P_{i}},E((B,K_{s},t_s), PU_{P_{i}}))$} 
$\xleftarrow{\makebox[6cm]{$PID_{MP_{i}},PID_{P_{i}},E((B,K_{s},t_s), PU_{P_{i}})$}}$\\
\begin{itemize}
\item Decrypts message and checks $t_s$ 
\item Retrieves $(B,K_{s})$ 
\item Computes $K^{\prime}_{s}$ = $B^x$mod$\alpha$ 
\item Checks whether $K^{\prime}_{s}$ ?= $K_{s}$ 
\end{itemize} & & \\
\hline
\end{tabular}
\end{table*}

\begin{figure*}[t]
\centering
\includegraphics[width=0.9\textwidth, height=.25\textheight]{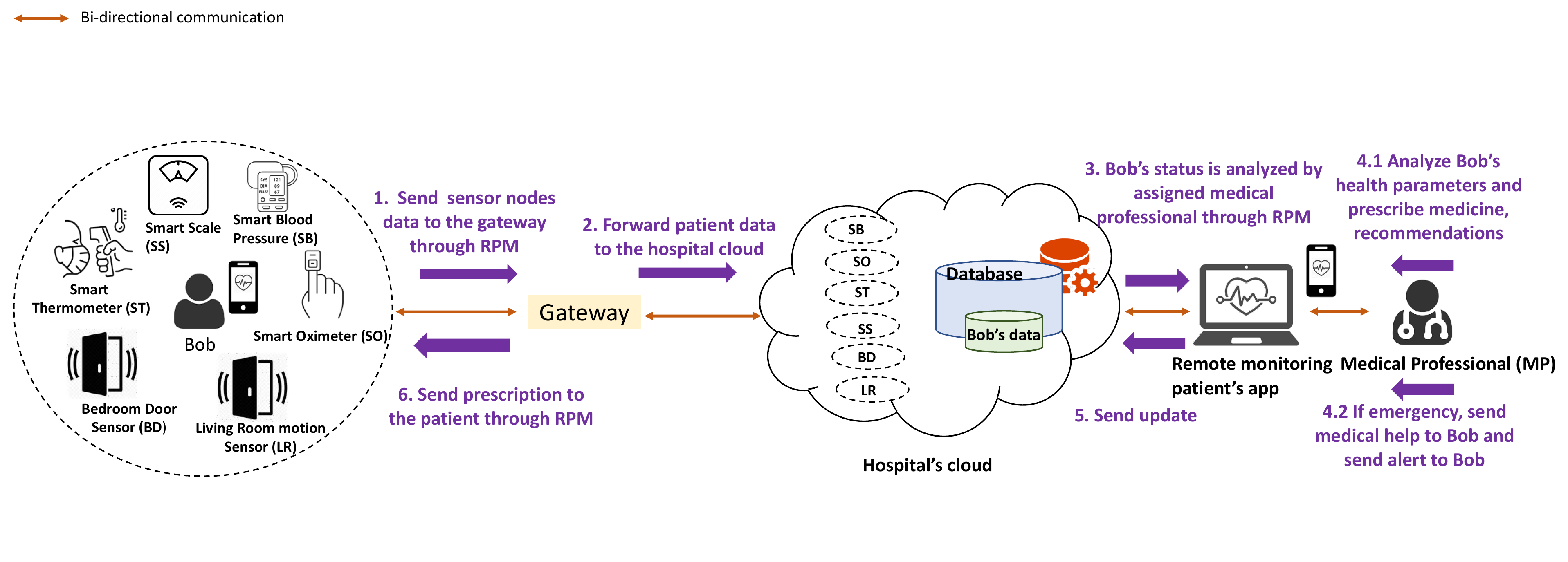}
\centering
\caption{Remote Patient Monitoring Use Case}
\label{fig:Usecase}
\end{figure*}

After successful verification, the patient $P_{i}$ sends a message that includes its pseudo ID $PID_{P_{i}}$, public key $PU_{P_{i}}$, pseudo ID of the medical professional $PID_{MP_{i}}$, and message generation timestamp $t_{s}$ to the medical professional $MP_{i}$ to authenticate himself/herself after successful verification. This message also contains some secret parameters to establish a secret shared key with the medical professional $PID_{MP_{i}}$ using Diffie-Hellman Key exchange protocol~\cite{diffie1976new} for secure RPM communication.

$P_{i}$ $\rightarrow$ $MP_{i}$ : $(PID_{P_{i}}$, $PID_{MP_{i}}$, $E(( A, \alpha , g$, ${PU_{P_{i}}}$, $t_{s})$, ${PU_{MP_{i}}}))$ 

In the above message, $\alpha$ is a large prime number, $g$ is a primitive root of $\alpha$, and $A$ $=$ $g^{x}$ $mod$ $\alpha$, where $x$ is a large random number private to the patient $P_{i}$. 

When the medical professional $MP_{i}$ receives the message, he/she decrypts the message using his/her private key ${PR_{P_{i}}}$. Then, verifies the message generation timestamp $t_{s}$. After that, he/she checks the the positive cuckoo filter $({PCF_{P}})$ and negative cuckoo filter $({NCF_{P}})$ to verify the authenticity of the patient $P_{i}$ using Algorithm~\ref{algo1} and the query result of the Table~\ref{table1}. Upon successful verification, the medical professional $MP_{i}$ calculates $B$ $=$ $g^{y}$ $mod$ $\alpha$ using the received parameters and here, $y$ is a large random number private to the patient $MP_{i}$. He/she also calculates the secret shared key $K_s$ $=$ $A^{y}$ $mod$ $\alpha$. Next, the medical professional $MP_{i}$ sends a message to the patient $P_{i}$ and the message contains his/her pseudo ID $PID_{MP_{i}}$, pseudo ID of the patient $PID_{P_{i}}$, calculated number $B$, secret shared key $K_s$, and message generation timestamp $t_{s}$.

$MP_{i}$ $\rightarrow$ $P_{i}$ : $(PID_{MP_{i}}$, $PID_{P_{i}}$, $E(( B, K_{s}$, $t_{s})$, ${PU_{P_{i}}}))$ 

Upon receiving the above message, the patient $P_{i}$ first checks the message generation timestamp $t_{s}$ to verify the freshness of the received message. Next, he/she computes the secret shared key $K_s^{'}$ $=$ $B^{x}$ $mod$ $\alpha$ using the received parameter $B$ and its secret number ${x}$. After that, the patient $P_{i}$ compares the computed secret shared key $K_s^{'}$ with the received secret shared key $K_s$. If two values are the same, then he/she uses this secret shared key $K_s$ to send and receive health-related information to the medical professional ${MP_{i}}$ for secure and fast communication in the RPM network. Table~\ref{M1} presents the proposed authentication process.

\subsection{Anomaly Detection of Sensor Nodes}
%\subsection{Misbehaving of Sensor Nodes}
The patient $P_{i}$ logins the account and sends his/her vital data captured by IoMT to the medical professional $MP_{i}$ by encrypting the message with the secret shared key $K_{s}$ established during the mutual authentication process. Gupta et al.~\cite{gupta2021detecting} proposed the anomaly detection model for the RPM ecosystem by analyzing the patient's behavior using both smart home and smart health devices. A traditional RPM ecosystem includes only IoMT, which makes it challenging to identify misbehaving sensor nodes. For instance, a smart blood pressure machine reports hypertension of 150, which is identified as an anomaly. However, it could be a false alarm due to various reasons, including faulty sensors, data poisoning, and intruders. In order to accurately identify anomalies, we develop the co-relations among smart home and smart health sensors.

\begin{algorithm}[!t]
%\SetAlgoLined
\caption{Anomaly Detection of sensor nodes $N$}
\label{algo2}
\begin{algorithmic}[1]
\STATE{Collect patient's the data $D_i$ from activated sensors $N$}
\STATE{Preparing the data $D_i$ by converting into numerical form}
\STATE{Normalize the data $D_i$}
\STATE{Create the sequences sets $(X^n,y^n)$ of data based on correlations}
\STATE{Take these input sequence sets $(X^n,y^n)$, where $n= 1,2,$\dots$,N$ from $N$ sensor nodes, initial model parameter $w_{z}$, local minibatch size $J$, number of local epochs $H$, learning rate $\alpha$, number of rounds $Q$, $h$ hidden layer.}

\STATE{Split local dataset $D_i$ to mini batches of size $J$ which are included into the set $J_i$ and fed horizontally to four LSTM cells.}

\FOR{each local epoch $j$ from 1 to $H$}
\FOR{batch $(X, y)$ $\in$ $J$}
\STATE${h\textsubscript{t} = LSTM(h\textsubscript{t-1}, x\textsubscript{t}, w\textsuperscript{n})}$
\STATE{$y$\textsuperscript{n} = ${\sigma(W\textsuperscript{FC}h\textsubscript{2nd}+Bias)}$}\\
\STATE{$u$\textsuperscript{n} = $w$\textsuperscript{n} - ${w_{z}^n}$ }\\
\STATE{$w_{z}^n$ $\leftarrow$ $w_{z}^n$  + $\frac{\alpha}{N}$ ${\mathlarger{\sum}}\textsubscript{$n$ $\in$ $D_i$} u\textsuperscript{n}$}
\ENDFOR
\ENDFOR 
\STATE{Update weights $w_{z}^n$ to hospital server and start training again until minimizing the error to build the anomaly detection model.}
\end{algorithmic}
\end{algorithm}

In this scenario, we consider a patient \textit{Bob} who is 34 years old and lives alone at his home and has been diagnosed with \textit{Obstructive Sleep Apnea (OSA)} disease. His assigned Medical Professional monitors him through RPM, shown in Fig. \ref{fig:Usecase}. His doctor prescribes medicine or treatment through the RPM ecosystem. In our work, we used this anomaly detection model~\cite{gupta2021detecting} to identify the misbehaving nodes in the RPM ecosystem using medical and smart home devices. For instance, a smart blood pressure machine reports hypertension of 150. However, other sensor nodes, such as the motion sensor, describe that the patient is moving normally in his living room. It can be easily identified that the smart blood pressure machine reports the wrong data of the patient. If any anomaly is identified using this anomaly detection model in the RPM ecosystem, that misbehaving report is sent to the RA. We provide Algorithm \ref{algo2}, showing the steps to identify the misbehaving sensor nodes for this model.

\section{Security Analysis and Formal Security Validation}
\label{security}
\begin{figure*}[b]
  \centering
  \begin{minipage}[b]{0.4\textwidth}
    \includegraphics[width=\textwidth]{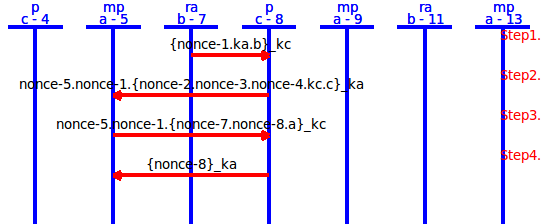}
    \caption{Message sequence chart of our proposed scheme using SPAN and AVISPA tools.}
    \label{simulation}
  \end{minipage}
  \hfill
  \begin{minipage}[b]{0.4\textwidth}
    \includegraphics[width=\textwidth]{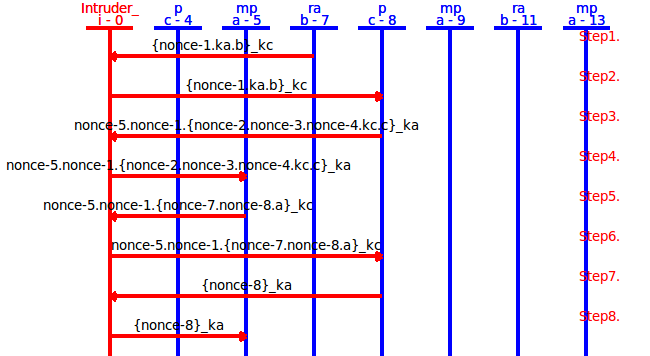}
    \caption{Message sequence chart of our proposed scheme in the presence of a malicious intruder.}
    \label{intruder}
  \end{minipage}
\end{figure*}

In this section, we analyze the security of our proposed authentication scheme. We also present a formal security validation using SPAN (Security Protocol ANimator)~\cite{SPAN} and AVISPA (Automated Validation of Internet Security Protocols and Applications)~\cite{Avispa} software tools.
\subsection{Security Analysis}
\subsubsection{Mutual authentication} In our proposed authentication scheme, the RA maintains the positive and negative cuckoo filters of medical professionals $(PCF_{MP}$, $NCF_{MP})$ and patients $(PCF_{P}$, $NCF_{P})$. RA signs the cuckoo filters and periodically broadcasts them. Both medical professional $MP_{i}$ and patient $P_{i}$ use the latest cuckoo filters to authenticate each other and establish a secret shared key for secure communication in the RPM network.
\subsubsection{Conditional privacy preservation} Upon successful registration, medical professionals $MP$ and patients $P$ get pseudo ID ($PID_{MP}$ for $MP$  and $PID_{P}$ for $P$ ) from the RA in our proposed scheme. Medical professionals and patients use pseudo IDs instead of real identities to protect their privacy. RA only knows the real identity of the medical professionals and patients. RA can also revoke the misbehaving medical professionals and patients from the RPM network in case of any disputes.     
\subsubsection{Resistance to replay attack} In our proposed authentication scheme, RA, medical professionals $MP$, and patients $P$ encrypt the message generation timestamp $t_{s}$ along with the message using the receiver's public key or the shared secret key between the sender and receiver. When the receiver receives the message, he/she decrypts the message and can check the freshness of the received message against the $t_{s}$. If $t_{s}$ is not fresh, the receiver rejects the received message to resist the replay attack. Our proposed scheme assumes that the clocks of RA, medical professionals $MP$, and patients $P$ are loosely synchronized.
\subsubsection{Resistance to message modification attack} The patient $P_{i}$  decrypts the message from the RA after a successful login process in the RPM network in our proposed scheme and gets the public key $PU_{MP_{i}}$ and pseud ID $PID_{MP_{i}}$ of the medical professional our proposed scheme. Next, the patient $P_{i}$ looks up into both the $PCF_{MP}$ and $NCF_{MP}$ to verify the received $PID_{MP_{i}}$. Upon successful verification, the patient $P_{i}$ encrypts its pseudo ID $PID_{P_{i}}$ and public key $PU_{P_{i}}$ along with some secret parameters using the public key $PU_{MP_{i}}$ of the medical professional $MP_{i}$. When the medical professional $MP_{i}$ receives the message, he/she decrypts the message and also verifies the authenticity of the patient's pseudo ID $PID_{P_{i}}$ using both the $PCF_{P}$ and $NCF_{P}$. After that, he/she generates secret parameters and computes the shared secret key using the received and generated secret parameters. Next, he/she sends the generated secret parameters and computed shared secret key to the patient $P_{i}$ by encrypting the message with the public key of the patient $PU_{P_{i}}$. Upon receiving the message from the medical professional $MP_{i}$, the patient $P_{i}$ recomputes the shared secret key using the received and generated secret parameters. If both the shared secret key values are the same, then the medical professional $MP_{i}$ is considered authentic. Otherwise, a message modification attack is detected.

\subsection{Formal Security Validation Using SPAN and AVISPA tools} We provide a formal validation of our proposed scheme using widely accepted SPAN and AVISPA tools in literature~\cite{wazid2017novel, moni2020efficient, yu2022robust,moni2022crease} to verify the security of our authentication protocol against replay attacks, man-in-the-middle attacks, and impersonation attacks.

We have three basic roles in our system model, namely $mp$ (Medical Professional), $ra$ (Registration Authority), and $p$ (Patient), which are represented by a, b, and c, respectively. In this model, $ka$, $kb$, and $kc$ denote the public key of a, b, and c. Each $mp$ and $p$ registers with the $ra$ and $ra$ initiates the start signal. After successful verification of the $p$, $ra$ sends the public key $ka$ and PID $nonce$-$\textit{1}$ of the $mp$ to the $p$. $p$ sends a message to the corresponding $mp$ that contains his/her PID $nonce$-$\textit{5}$ and secret parameters ($nonce$-$\textit{2}$, $nonce$-$\textit{3}$, and $nonce$-$\textit{4}$) to establish a secret shared key $nonce$-$\textit{8}$ between $p$ and $mp$. Fig.~\ref{simulation} shows the message sequence chart of our proposed scheme using SPAN and AVISPA tools.

The message sequence chart for malicious intruder $i$ using SPAN and AVISPA tools is presented in Fig.~\ref{intruder}. The implementation result shows that the malicious intruder $i$ can only listen to and forward messages. However, he/she is unable to read or modify the messages. The message sequences in the presence of $i$ in our proposed authentication scheme are described as follows:

\begin{itemize}
\item \textit{Step1:} The $ra$ sends the PID $nonce$-$\textit{1}$ and public key $ka$ of $mp$ by encrypting with public key of $p$ $kc$ to $p$.
\item \textit{Step 2:} The malicious intruder $i$ listens to the message and forwards it to the $p$.
\item \textit{Step 3:} The $p$ encrypts his/her pubic key $kc$ and some secret parameters ($nonce$-$\textit{2}$, $nonce$-$\textit{3}$, and $nonce$-$\textit{4}$) with $ka$ and sends to the $mp$ along with its PID $nonce$-$\textit{5}$.
\item \textit{Step 4:} Since the secret parameters are encrypted with $ka$, the malicious intruder $i$ is unable to read the message.
\item\textit{Step 5:} The $mp$ decrypts the message using its private key and generates the secret shared key $nonce$-$\textit{8}$. Next, $mp$ sends the $nonce$-$\textit{8}$ and a secret parameter $nonce$-$\textit{7}$ by encrypting with $kc$ to the $p$.  
\item \textit{Step 6:} The malicious intruder $i$ is unable to read or modify the message, he/she only forwards the message.
\item\textit{Step 7:} After receiving the message, $p$ recalculates the secret shared key $nonce$-$\textit{8}$ and sends it back to the $mp$ as an acknowledgment message. 
\item \textit{Step 8:} The malicious intruder $i$ only listens to the message as it is decrypted with the $ka$ and forwards to the $mp$.
\end{itemize}

\section{Conclusion and Future Work}
\label{conclusion}
In this paper, we propose a secure and efficient privacy-preserving authentication scheme based on the cuckoo filter for the RPM network. Our scheme leverages cuckoo filters, making it efficient and lightweight for medical professionals and patients to authenticate each other and establish a secret shared key for secure communication in the RPM network. An anomaly detection model is also provided to identify misbehaving sensor nodes in the RMP network. This proposed authentication scheme mitigates various risks, including message modification attacks, replay attacks, and man-in-the-middle attacks between patients and medical professionals in the RPM network. Our future work entails test-bed setup and implementation of our approach, along with its performance evaluation to enhance the security of the RPM network. 
{
\bibliographystyle{IEEEtran}
\bibliography{References}

\begin{thebibliography}{10}
\providecommand{\url}[1]{#1}
\csname url@samestyle\endcsname
\providecommand{\newblock}{\relax}
\providecommand{\bibinfo}[2]{#2}
\providecommand{\BIBentrySTDinterwordspacing}{\spaceskip=0pt\relax}
\providecommand{\BIBentryALTinterwordstretchfactor}{4}
\providecommand{\BIBentryALTinterwordspacing}{\spaceskip=\fontdimen2\font plus
\BIBentryALTinterwordstretchfactor\fontdimen3\font minus
  \fontdimen4\font\relax}
\providecommand{\BIBforeignlanguage}[2]{{%
\expandafter\ifx\csname l@#1\endcsname\relax
\typeout{** WARNING: IEEEtran.bst: No hyphenation pattern has been}%
\typeout{** loaded for the language `#1'. Using the pattern for}%
\typeout{** the default language instead.}%
\else
\language=\csname l@#1\endcsname
\fi
#2}}
\providecommand{\BIBdecl}{\relax}
\BIBdecl

\bibitem{hayajneh2016secure}
T.~Hayajneh, B.~J. Mohd, M.~Imran, G.~Almashaqbeh, and A.~V. Vasilakos,
  ``Secure authentication for remote patient monitoring with wireless medical
  sensor networks,'' \emph{Sensors}, vol.~16, no.~4, p. 424, 2016.

\bibitem{alzahrani2020secure}
B.~A. Alzahrani, A.~Irshad, K.~Alsubhi, and A.~Albeshri, ``A secure and
  efficient remote patient-monitoring authentication protocol for cloud-iot,''
  \emph{International Journal of Communication Systems}, vol.~33, no.~11, p.
  e4423, 2020.

\bibitem{soni2021privacy}
M.~Soni and D.~K. Singh, ``Privacy-preserving authentication and key-management
  protocol for health information systems,'' in \emph{Data Protection and
  Privacy in Healthcare}.\hskip 1em plus 0.5em minus 0.4em\relax CRC Press,
  2021, pp. 37--50.

\bibitem{shuai2020efficient}
M.~Shuai, B.~Liu, N.~Yu, L.~Xiong, and C.~Wang, ``Efficient and
  privacy-preserving authentication scheme for wireless body area networks,''
  \emph{Journal of Information Security and Applications}, vol.~52, p. 102499,
  2020.

\bibitem{cawthra1800securing}
J.~Cawthra, N.~Grayson, R.~Pulivarti, B.~Hodges, J.~Kuruvilla, K.~Littlefield,
  J.~Snyder, S.~Wang, R.~Williams, K.~Zheng \emph{et~al.}, ``Securing
  telehealth remote patient monitoring ecosystem,'' \emph{NIST SPECIAL
  PUBLICATION}, p. 30B, 1800.

\bibitem{soleymani2022privacy}
S.~A. Soleymani, S.~Goudarzi, M.~H. Anisi, A.~Jindal, N.~Kama, and S.~A.
  Ismail, ``A privacy-preserving authentication scheme for real-time medical
  monitoring systems,'' \emph{IEEE Journal of Biomedical and Health
  Informatics}, 2022.

\bibitem{soni2021laka}
M.~Soni and D.~K. Singh, ``Laka: lightweight authentication and key agreement
  protocol for internet of things based wireless body area network,''
  \emph{Wireless Personal Communications}, pp. 1--18, 2021.

\bibitem{ali2020securing}
Z.~Ali, S.~A. Chaudhry, M.~S. Ramzan, and F.~Al-Turjman, ``Securing smart city
  surveillance: A lightweight authentication mechanism for unmanned vehicles,''
  \emph{IEEE Access}, vol.~8, pp. 43\,711--43\,724, 2020.

\bibitem{shahidinejad2021light}
A.~Shahidinejad, M.~Ghobaei-Arani, A.~Souri, M.~Shojafar, and S.~Kumari,
  ``Light-edge: a lightweight authentication protocol for iot devices in an
  edge-cloud environment,'' \emph{IEEE consumer electronics magazine}, vol.~11,
  no.~2, pp. 57--63, 2021.

\bibitem{moni2022lightweight}
S.~S. Moni and D.~Manivannan, ``{A lightweight privacy-preserving V2I mutual
  authentication scheme using Cuckoo filter in VANETs},'' in \emph{2022 IEEE
  19th Annual Consumer Communications \& Networking Conference (CCNC)}.\hskip
  1em plus 0.5em minus 0.4em\relax IEEE, 2022, pp. 815--820.

\bibitem{grashofer2018towards}
J.~Grash{\"o}fer, F.~Jacob, and H.~Hartenstein, ``Towards application of cuckoo
  filters in network security monitoring,'' in \emph{2018 14th International
  Conference on Network and Service Management (CNSM)}.\hskip 1em plus 0.5em
  minus 0.4em\relax IEEE, 2018, pp. 373--377.

\bibitem{yang2021blockchain}
X.~Yang, X.~Yang, X.~Yi, I.~Khalil, X.~Zhou, D.~He, X.~Huang, and S.~Nepal,
  ``Blockchain-based secure and lightweight authentication for internet of
  things,'' \emph{IEEE Internet of Things Journal}, vol.~9, no.~5, pp.
  3321--3332, 2021.

\bibitem{gupta2020access}
D.~Gupta, S.~Bhatt, M.~Gupta, O.~Kayode, and A.~S. Tosun, ``Access control
  model for google cloud iot,'' in \emph{2020 IEEE 6th Intl conference on big
  data security on cloud (BigDataSecurity), IEEE Intl conference on high
  performance and smart computing,(HPSC) and IEEE Intl conference on
  intelligent data and security (IDS)}.\hskip 1em plus 0.5em minus 0.4em\relax
  IEEE, 2020, pp. 198--208.

\bibitem{gupta2020learner}
D.~Gupta, O.~Kayode, S.~Bhatt, M.~Gupta, and A.~S. Tosun, ``Learner’s
  dilemma: Iot devices training strategies in collaborative deep learning,'' in
  \emph{2020 IEEE 6th World Forum on Internet of Things (WF-IoT)}.\hskip 1em
  plus 0.5em minus 0.4em\relax IEEE, 2020, pp. 1--6.

\bibitem{moni2020scalable}
S.~S. Moni and D.~Manivannan, ``A scalable and distributed architecture for
  secure and privacy-preserving authentication and message dissemination in
  {VANETs},'' \emph{Internet of Things}, vol.~13, p. 100350, 2020.

\bibitem{gupta2021game}
D.~Gupta, S.~Bhatt, P.~Bhatt, M.~Gupta, and A.~S. Tosun, ``Game theory based
  privacy preserving approach for collaborative deep learning in iot,'' in
  \emph{Deep Learning for Security and Privacy Preservation in IoT}.\hskip 1em
  plus 0.5em minus 0.4em\relax Springer, 2021, pp. 127--149.

\bibitem{gupta2021future}
D.~Gupta, S.~Bhatt, M.~Gupta, and A.~S. Tosun, ``Future smart connected
  communities to fight covid-19 outbreak,'' \emph{Internet of Things}, vol.~13,
  p. 100342, 2021.

\bibitem{aslan2021intelligent}
{\"O}.~Aslan, M.~Ozkan-Okay, and D.~Gupta, ``Intelligent behavior-based malware
  detection system on cloud computing environment,'' \emph{IEEE Access},
  vol.~9, pp. 83\,252--83\,271, 2021.

\bibitem{ozkan2021comprehensive}
M.~Ozkan-Okay, R.~Samet, {\"O}.~Aslan, and D.~Gupta, ``A comprehensive
  systematic literature review on intrusion detection systems,'' \emph{IEEE
  Access}, 2021.

\bibitem{gupta2021detecting}
D.~Gupta, M.~Gupta, S.~Bhatt, and A.~S. Tosun, ``Detecting anomalous user
  behavior in remote patient monitoring,'' in \emph{2021 IEEE 22nd
  International Conference on Information Reuse and Integration for Data
  Science (IRI)}.\hskip 1em plus 0.5em minus 0.4em\relax IEEE, 2021, pp.
  33--40.

\bibitem{gupta2021hierarchical}
D.~Gupta, O.~Kayode, S.~Bhatt, M.~Gupta, and A.~S. Tosun, ``Hierarchical
  federated learning based anomaly detection using digital twins for smart
  healthcare,'' in \emph{2021 IEEE 7th International Conference on
  Collaboration and Internet Computing (CIC)}.\hskip 1em plus 0.5em minus
  0.4em\relax IEEE, 2021, pp. 16--25.

\bibitem{fan2014cuckoo}
B.~Fan, D.~G. Andersen, M.~Kaminsky, and M.~D. Mitzenmacher, ``Cuckoo filter:
  Practically better than bloom,'' in \emph{Proceedings of the 10th ACM
  International on Conference on emerging Networking Experiments and
  Technologies}, 2014, pp. 75--88.

\bibitem{diffie1976new}
W.~Diffie and M.~E. Hellman, ``New directions in cryptography,'' \emph{IEEE
  Transaction on Information Theory}, vol.~22, no.~6, 1976.

\bibitem{SPAN}
Irisa, ``{SPAN},'' \url{http://people.irisa.fr/Thomas.Genet/span/}, 2006.

\bibitem{Avispa}
Avispa, ``{AVISPA},'' \url{http://www.avispa‐project.org}, 2002.

\bibitem{wazid2017novel}
M.~Wazid, A.~K. Das, N.~Kumar, M.~Conti, and A.~V. Vasilakos, ``A novel
  authentication and key agreement scheme for implantable medical devices
  deployment,'' \emph{IEEE journal of biomedical and health informatics},
  vol.~22, no.~4, pp. 1299--1309, 2017.

\bibitem{moni2020efficient}
S.~S. Moni and D.~Manivannan, ``An efficient {RSU} authentication scheme based
  on {Merkle Hash Tree for VANETs},'' in \emph{ICC 2020-2020 IEEE International
  Conference on Communications (ICC)}.\hskip 1em plus 0.5em minus 0.4em\relax
  IEEE, 2020, pp. 1--7.

\bibitem{yu2022robust}
S.~Yu and Y.~Park, ``A robust authentication protocol for wireless medical
  sensor networks using blockchain and physically unclonable functions,''
  \emph{IEEE Internet of Things Journal}, 2022.

\bibitem{moni2022crease}
S.~S. Moni and D.~Manivannan, ``{CREASE:} {C}ertificateless and
  {REused}-pseudonym based {Authentication Scheme for Enabling security and
  privacy in VANETs},'' \emph{Internet of Things}, p. 100605, 2022.

\end{thebibliography}
}
\end{document}